\documentstyle[epsfig,indentfirst,bm,12pt]{article}
\begin{document}
\title{\Large{\bf{The uncertainties due to quark energy loss on  determining  nuclear sea quark distribution
from nuclear Drell-Yan data}}}
\author{C.G.  Duan$^{1,2,6}$ \footnote{\tt{ E-mail:duancg$@$mail.hebtu.edu.cn}},
 N.  Liu$^{1,3}$, G.L.Li$^{4,5,6}$}

\date{}

\maketitle \noindent {\small 1.Department of Physics, Hebei Normal
               University, Shijiazhuang 050016,P.R.China}\\
{\small 2.Hebei Advanced Thin Films Laboratory, shijiazhuang 050016,P.R.China}\\
{\small 3.College of Mathematics and Physics, Shijiazhuang University of Economics, Shijiazhuang 050031,P.R. China}\\
{\small 4.Department of Physics, Nanjing Xiaozhuang University,
Nanjing ,210017,P.R.China}\\
 {\small 5.Institute of high energy physics
,The Chinese academy of sciences ,Beijing,100039,P.R.China}\\
{\small 6.CCAST(World Laboratory), P.O.Box8730, Beijing
          100080,P.R.China}

\baselineskip 9mm \vskip 0.5cm
\begin{abstract}

By means of two different parametrizations of  quark energy loss and
the nuclear parton distributions determined only with lepton-nuclear
deep inelastic scattering experimental data,  a leading order
phenomenological analysis is performed on the nuclear Drell-Yan
differential cross section ratios as a function of the quark
momentum fraction in the beam proton and target nuclei for E772
experimental data. It is shown that there is the quark energy loss
effect in nuclear Drell-Yan process apart from the nuclear effects
on the parton distribution as in deep inelastic scattering. The
uncertainties due to quark energy loss effect is quantified on
determining nuclear sea quark distribution by using nuclear
Drell-Yan data. It is found that the quark energy loss effect on
nuclear Drell-Yan cross section ratios make greater with the
increase of quark momentum fraction in the target nuclei.  The
uncertainties from quark energy loss become bigger as  the nucleus A
come to be heavier.  The Drell-Yan data on proton incident middle
and heavy nuclei versus deuterium would result in  an overestimate
for nuclear modifications on sea quark distribution functions with
neglecting the quark energy loss. Our results are hoped to provide
good directional information on the magnitude and form of nuclear
modifications on sea quark distribution functions by means of the
nuclear Drell-Yan experimental data.



\end{abstract}

\newpage
\vskip 0.5cm

\noindent {\bf 1 Introduction }

\hspace{0.5cm} The nuclear quark and gluon distributions have been
one of the most active frontiers in nuclear physics because
universal process-independent nuclear parton distribution functions
are a key element in computing differential cross sections from
nuclear collisions. The nuclear parton distributions directly affect
the interpretation of the data collected from the high energy
nuclear reactions$^{[1,2]}$, especially at the Relativistic Heavy
Ion Collider (RHIC) and the Large Hadron Collider (LHC).  The
precise nuclear parton distribution functions are also very
important in finding the new physics phenomena and determining the
electro-weak parameters, neutrino masses and mixing angles in
neutrino physics.

\hspace{0.5cm} After the discovery of the EMC effect$^{[3]}$, it is
well known that nuclear  parton distribution functions are mutually
different from those in free nucleon.  The nuclear modifications
relative to the nucleon parton distribution functions, are usually
referred to as the nuclear effects on the parton distribution
functions, which include nuclear shadowing, anti-shadowing, EMC
effect and Fermi motion effect in different regions of parton
momentum fraction.  The origin of the nuclear effects is still under
debate in theory, and it is considered that different mechanisms are
responsible for the effects in the different regions of parton
momentum fraction$^{[4]}$. Up to now, almost all of the data on
nuclear dependence is from charged lepton deep inelastic scattering
experiments, which are sensitive to the charge-weighted sum of all
quark and anti-quark distributions. From the charged lepton deep
inelastic scattering off nuclei, the nuclear valence quark
distributions are relatively well determined in the medium and large
Bjorken $x$ regions. However, the charged lepton deep inelastic
scattering would not be sensitive to the nuclear sea quark
distributions. In order to pin down the nuclear anti-quark
distributions, it is desirable that the nuclear Drell-Yan
reaction$^{[5]}$ is an ideal complementary tool in proton-nucleus
collisions.

\hspace{0.5cm} The nuclear Drell-Yan process  is induced by the
annihilation of a quark(anti-quark) with a target anti-quark(quark)
into a virtual photon which subsequently decays into a pair of
oppositely-charged lepton. Therefore, the nuclear Drell-Yan process
is closely related to the quark distribution functions in target
nuclei. It is naturally expected that the nuclear Drell-Yan
reaction, which is a complementary tool to probe the structure of
nuclei in lepton-nucleus deep inelastic scattering, can be used to
extract the sea quark distributions in the target nuclei. However,
in high energy proton-nucleus scattering, the projectile rarely
retains a major fraction of its momentum in traversing the nucleus.
The quark and gluon in the induced proton can loss a finite fraction
of its energy  due to the  multiple collisions and repeated energy
loss in the nuclear target. In the view, the initial-state
interactions are very important in nuclear Drell-Yan process since
the dimuon in the final state does not interact strongly with the
partons in the nuclei. The quark energy loss effect in  nuclear
Drell-Yan process is another nuclear effect apart from the nuclear
effects on the parton distribution as in deep inelastic scattering.

\hspace{0.5cm}  In consideration of the strong necessities of
precise nuclear parton distributions, the global analysis of nuclear
parton distribution functions have been proposed in the recent
years. So far, three groups have presented global  analyses of the
nuclear parton distribution functions analogous to those of the free
proton. These are the ones by Eskola et al. ( EKS98 $^{[6-7]}$ and
EKPS$^{[8]}$), by Hirai et al. (HKM$^{[9]}$, HKN04$^{[10]}$ and
HKN07$^{[11]}$), and by de Florian and Sassot nDS $^{[12]}$.  It is
noticeable that EKPS and HKN employed Fermilab E772$^{[13]}$ and
E866 $^{[14]}$ nuclear Drell-Yan  data, EKS98 and nDS included E772
experimental data,  and HKM proposed the nuclear parton
distributions which were determined by means of the existing
experimental data on nuclear structure functions without including
the proton-nucleus Drell-Yan process.

\hspace{0.5cm}In nuclear Drell-Yan process, the quark energy loss
effect  would lead to a degradation of the quark momentum prior to
annihilation, further resulting in a less energetic dimuon.
Therefore,  the quark energy loss effect can drop the differential
cross sections as a function of the quark momentum fraction in the
beam proton and target nuclei. We have investigated the nuclear
Drell-Yan differential cross section ratios as the function of quark
momentum fraction of the beam proton in the framework of Glauber
model and two typical kinds of quark energy loss parametrization
$^{[15-17]}$. It is proved that there is quark energy loss effect in
nuclear Drell-Yan reactions, which is an ideal tool to study the
energy loss of the fast quark moving through cold nuclei. In the
recent paper$^{[18]}$,  a next-to-leading order and a leading order
analysis are performed of the differential cross section ratios from
the proton-induced Drell-Yan reaction off nuclei. It is found that
the next-to-leading order corrections can be negligible on the
differential cross section ratios for the current Fermilab and
future lower proton energy. The Fermilab E866 experimental data were
used in these works.

\hspace{0.5cm}The nuclear Drell-Yan differential cross section
ratios as a function of the quark momentum fraction in target nuclei
are currently used to pin down the nuclear sea quark distribution
functions. However, the quark energy loss effect can suppress the
differential cross sections versus the quark momentum fraction.  In
order to obtained the precise nuclear sea quark distributions, the
uncertainties due to quark energy loss should be carefully
researched on determining nuclear sea quark distribution from
Drell-Yan reaction off nuclei, which is the main purpose of this
present paper. The Fermilab E866 measured the 800GeV proton incident
Drell-Yan cross section ratios of per nucleon for Fe and W nuclei
over Be. The impact of quark energy loss is canceled partly out in
the Drell-Yan cross section ratios. The Fermilab E772 presented the
Drell-Yan cross section ratios of various nuclei (C, Ca, Fe and W)
versus deuterium in same energetic proton beam. Because the small
nuclear effects in deuterium are neglected, the E772 experimental
data help us to probe the quark energy loss effect in a more
excellent way. In this paper, the studies on quark energy loss are
extended  to the E772 data. The influence of quark energy loss is
quantified on the nuclear Drell-Yan differential cross section
ratios. The uncertainties due to quark energy loss is elucidated on
determining nuclear sea quark distribution from nuclear Drell-Yan
process.

\hspace{0.5cm} The paper is organized as follows. In sect.2, a brief
formalism for the Drell-Yan differential cross section and two
different parametrizations of of quark energy loss are presented.
The section 3 contains our results and analysis about the
uncertainties due to quark energy loss on determining nuclear sea
quark distributions. The summary is given in sect.4.

\vskip 0.5cm

{\bf 2 The formalism for Drell-Yan differential cross section }

\hspace{0.5cm} In the Drell-Yan process$^{[4]}$, the leading-order
contribution is quark-antiquark annihilation into a lepton pair. The
annihilation cross section can be obtained from the
$q\bar{q}\rightarrow\l^{+}l^{-}$ cross section, which is
\begin{equation}
  \sigma [q\bar{q}\rightarrow\l^{+}l^{-}]=\frac{4\pi\alpha_{em}^2}{9M^2}e^2_f,
\end{equation}
where $\alpha_{em}$ is the fine-structure constant, $e_f$ is the
charge of the quark, and  $M$ is the invariant mass of the lepton
pair. The nuclear Drell-Yan differential cross section can be
written as
\begin{equation}
 \frac{d^2\sigma}{dx_1dx_2}=\frac{4\pi\alpha_{em}^2}{9sx_1x_2}
 \sum_{f}e^2_f[q^p_f(x_1,Q^2)\bar{q}^A_f(x_2,Q^2)
 +\bar{q}^p_f(x_1,Q^2)q^A_f(x_2,Q^2)],
\end{equation}
where$\sqrt{s}$ is the center of mass energy of the hadronic
collision,  $x_1$ and $x_2$ is the momentum fraction of the partons
in the beam and target respectively, the sum is carried out over the
light flavor $f=u,d,s$, and $q^{p(A)}_{f}(x,Q^2)$ and ${\bar
q}^{p(A)}_{f}(x,Q^2)$ are the quark and anti-quark distributions in
the proton (nucleon in the nucleus A).

\hspace{0.5cm}The energy loss of fast partons in nuclei have been
 studied by Gavin and Milana $^{[19]}$, Brodsky and
Hoyer$^{[20]}$, and by Baier et al.$^{[21]}$ respectively. Two
typical kinds of quark energy loss expressions would be introduced
with basing on the theoretical researches.  One is rewritten as
\begin{equation}
\Delta x_1= {\alpha}\frac{<L>_A}{E_p},
\end{equation}
where $\alpha$ denotes the  incident quark energy loss per unit
length in nuclear matter,  $<L>_A=3/4(1.2A^{1/3)}$fm$^{[22]}$ is the
average path length of the incident quark in the nucleus A, and
$E_p$ is the energy of the incident proton.  In addition to the
linear quark energy loss rate, another one is presented as
\begin{equation}
\Delta x_1= {\beta}\frac{<L>^2_A}{E_p}.
\end{equation}
Obviously, the quark energy loss is quadratic with the path length.
In what follows, the two different parametrizations  are named to
the linear and quadratic quark energy loss respectively. The quark
energy loss in nuclei shifts the incident quark momentum fraction
from $x'_1=x_1+\Delta x_1$ to $x_1$ at the point of fusion. After
considering the  quark energy loss in nuclei, the nuclear Drell-Yan
differential cross section can be expressed as
\begin{equation}
 \frac{d^2\sigma}{dx_1dx_2}=\frac{4\pi\alpha_{em}^2}{9sx_1x_2}
 \sum_{f}e^2_f[q^p_f(x'_1,Q^2)\bar{q}^A_f(x_2,Q^2)
 +\bar{q}^p_f(x'_1,Q^2)q^A_f(x_2,Q^2)].
\end{equation}
With calculating the integral of the differential cross section
above, the nuclear Drell-Yan production cross section is given by
\begin{equation}
 \frac{d\sigma}{dx_{1(2)}}=\int dx_{2(1)}\frac{d^2\sigma}{dx_1dx_2}.
\end{equation}
The integral range is determined according to the relative
experimental kinematic region.

\vskip 0.5cm

{\bf 3 Results  and discussion }

\hspace{0.5cm}In the nuclear Drell-Yan experiments,  the ratios are
measured of  Drell-Yan cross sections on two different nuclear
targets bombarded by proton,
\begin{equation}
R_{A_{1}/A_{2}}(x_{1(2)})=\frac{d\sigma^{p-A_{1}}}{dx_ {1(2)}}
/{\frac {d\sigma^{p-A_{2}}}{dx_{1(2)}}}.
\end{equation}
The available nuclear Drell-Yan data are in the form of ratios over
deuterium and beryllium, $R_{A/Be}(x_{1(2)})$ for the Fermilab
Experiment866, and $R_{A/D}(x_{1(2)})$ for the Fermilab
Experiment772, respectively. The $x_1$ dependence of cross section
rations $R_{A_{1}/A_{2}}(x_{1})$ provide the best measure of the
energy loss of the incident quarks in cold nuclear matter. The $x_2$
dependence of cross section rations $R_{A_{1}/A_{2}}(x_{2})$ are
used to complementarily obtain the nuclear sea quark distributions.
In this work, by combining HKM cubic type of nuclear parton
distribution with the quark energy loss, the global $\chi^2$
analysis to the E772 experimental data are performed in the
perturbative QCD leading order.   $R_{A/D}(x_{1})$ and
$R_{A/D}(x_{2})$ are respectively calculated and  compared with the
E772 experimental data on  nuclear Drell-Yan differential cross
section rations.

\hspace{0.5cm}For the nuclear Drell-Yan differential cross section
rations $R_{A/D}(x_{1})$,  the obtained $\chi^2$ value is
$\chi^2=234.36$ for the 122 total data points,  The $\chi^2$ per
degrees of freedom is  $\chi^2/d.o.f.=1.92$ without quark energy
loss effect. As for the rations $R_{A/D}(x_{2})$,  $\chi^2$ value is
$\chi^2=176.39$ for the 36 total data points, the $\chi^2$ per
degrees of freedom is given by $\chi^2/d.o.f.= 4.89$. It is apparent
that theoretical results without energy loss effect deviate indeed
from  the E772 experimental data.  After adding the fast quark
energy loss effect on the ratios $R_{A/D}(x_{1})$ and
$R_{A/D}(x_{2})$, the $\chi^2$ per degrees of freedom, $\alpha$ and
$\beta$ are summarized in Table 1. for the linear and quadratic
quark energy loss formula.
\tabcolsep0.5cm
\begin{table}
\caption{The list of $\chi^2/d.o.f.$, $\alpha$ and $\beta$ by
fitting the E772 experimental data.}
\begin{center}
\begin{tabular}{ccccc}\hline
                   &  Exp. data       & $\chi^2/d.o.f.(\alpha)$ & $ \chi^2/d.o.f.(\beta)$  \\\hline
                   &$R_{A/D}(x_{1})$  & 1.33 (1.26)             & 1.35(0.23)         \\
                   &$R_{A/D}(x_{2})$  &  1.10 (1.31)            & 1.53 (0.23)         \\\hline
\end{tabular}
\end{center}
\end{table}
It can be found that the theoretical results with energy loss effect
are in good agreement with the experimental data. It is concluded
that there is the quark energy loss effect apart from the nuclear
effects on the parton distribution as in deep inelastic scattering.
Meanwhile, the calculated $\chi^2/d.o.f.$ given by the linear quark
energy loss is nearly the same as that from the quadratic quark
energy loss for $R_{A/D}(x_{1})$. The small differentness of
$\alpha(1.26, 1.31)$ and same $\beta(0.23)$ value are obtained by
fitting $R_{A/D}(x_{1})$ and $R_{A/D}(x_{2})$ data, which is
different from the results with E866 data$^{[18]}$. In fact, the
values of the parameter $\alpha$ (or $\beta$) in the quark energy
loss expression should be the same for fitting the ratios
$R_{A_{1}/A_{2}}(x_{1})$ or $R_{A_{1}/A_{2}}(x_{2})$ from the
nuclear Drell-Yan experiment if the experimental data are
sufficiently precise. Therefore, it is expected that the good
$\alpha$ and  $\beta$ are determined from future precise nuclear
Drell-Yan experiment. As an example, the calculated results with
linear energy loss expression are shown in Fig.1 and Fig.2 against
the E772 experimental data, which is the nuclear Drell-Yan
differential cross section ratios for Ca to D and W to D as
functions of $x_1$ for various interval of $M$, respectively. The
solid curves are the ratios with only the nuclear effects on the
parton distribution. The dotted curves correspond to the results
from an linear quark energy loss  with nuclear effect on structure
function. In Fig.3,
 the computed ratios $R_{A/D}(x_{2})$ are compared with the E772 data
  for C to D, Ca to D, Fe to D and W over D as
functions of $x_2$, respectively. Apart from the same comments with
Fig.1 and Fig.2, the dash curves stand for the results from the
quadratic energy loss.

\hspace{0.5cm}The calculated results above demonstrate that the
quark energy loss effect suppress obviously the nuclear Drell-Yan
differential cross section ratios versus the quark momentum fraction
in target nuclei. In order to analyze the uncertainties due to quark
energy loss on  determining  nuclear sea quark distribution, we
quantify the quark energy loss effect on the nuclear Drell-Yan
differential cross section ratios.  The ratios $RR_{A/D}(x_{2})$ on
$R_{A/D}(x_{2})$ without quark energy loss to those with linear
quark energy loss $(\alpha=1.26) $  are calculated and tabulated in
Table 2. The similar results can be obtained for the quadratic quark
energy loss. The E772 and E886 cover respectively the ranges
$0.1\leq x_2\leq0.3$ and $0.01\leq x_2\leq0.12$. For completeness,
the ratios $RR_{A/Be}(x_{2})$ are given for the E866 experimental
data$^{[18]}$in Table 3.  It is shown that  the suppression due to
quark energy loss are approximately $2\%$ to $3\%$ for
$R_{Fe/Be}(x_{2})$ and $4\%$ to $5\%$ for $R_{W/Be}(x_{2})$ in the
ranges  $0.03\leq x_2\leq0.12$, respectively. As for the E772
experimental data in the ranges $0.1\leq x_2\leq0.3$,  the
variations from quark energy loss are roughly $1\%$ to $4\%$ for
$R_{C/D}(x_{2})$, $2\%$ to $8\%$ for $R_{Ca/D}(x_{2})$, $3\%$ to
$9\%$ for $R_{Fe/D}(x_{2})$, and $4\%$ to $16\%$ for
$R_{W/D}(x_{2})$, respectively. It is found that the quark energy
loss is canceled partly out in the proton-induced Drell-Yan cross
section ratios for Fe to Be and W to Be from E866 data. The quark
energy loss effect on $R_{A_1/A_2}(x_{2})$ make greater with the
increase of momentum fraction of the target parton. It is indicated
that the heavier the target nuclei A, the bigger the uncertainties
due to quark energy loss. The quark energy loss effect do not drop
the nuclear Drell-Yan cross section in the same ratio. By taking
account of the small energy loss effect, the $R_{C/D}(x_{2})$ data
in the region $ x_2<0.1$ can be employed safely to determine the
nuclear sea quark distribution functions. However, the  remaining
ones  are not remarkably suitable for the constraints of the nuclear
anti-quark distribution. The large degradation would result in an
overestimate for nuclear modifications on sea quark distribution
functions if the quark energy loss effect is not put in our mind.
The suppression from energy loss would change not only the magnitude
but also the form of nuclear modifications on sea quark distribution
functions by means of the nuclear Drell-Yan experimental data
because of the different compression ratio. We hope that our results
from $RR_{A_1/A_2}(x_{2})$ can provide  good directional information
on the  nuclear modifications on sea quark distribution functions.
 \tabcolsep0.5cm
\begin{table}
\caption{The ratios of $R_{A/D}(x_{2})$ without quark energy loss to
those with linear quark energy loss$(\alpha=1.26) $.}
\begin{center}
\begin{tabular}{ccccccc}\hline
                   &{$x_2$}   & 0.04 & 0.10 & 0.18 & 0.24 & 0.30 \\\hline
                   &$RR_{C/D}(x_{2})$      & 1.010 & 1.014 &1.023  & 1.030  & 1.037 \\
                   &$RR_{Ca/D}(x_{2})$ &      1.022 & 1.030 & 1.049& 1.064 & 1.078\\
                   &$RR_{Fe/D}(x_{2})$      & 1.026  & 1.036 & 1.058  & 1.076  & 1.093 \\
                   &$RR_{W/D}(x_{2})$ &      1.045 & 1.062 & 1.100& 1.131 & 1.161\\ \hline
\end{tabular}
\end{center}
\end{table}
\tabcolsep0.5cm
\begin{table}
\caption{The ratios $RR_{A/Be}(x_{2})$ from the E866 experimental
data$^{[18]}$ with linear quark energy loss$(\alpha=1.27) $.}
\begin{center}
\begin{tabular}{ccccccc}\hline
                   &$x_2$   & 0.03 & 0.05 & 0.07 & 0.09 & 0.12 \\\hline
                   &$RR_{Fe/Be}(x_{2})$ & 1.018  & 1.017 &1.019  & 1.022  & 1.027 \\
                   &$RR_{W/Be}(x_{2})$ &      1.038 & 1.036 & 1.040& 1.045 & 1.055\\ \hline
\end{tabular}
\end{center}
\end{table}

\hspace{0.5cm}Let us now discuss the impact of the nuclear Drell-Yan
experimental data on the global analysis of nuclear parton
distribution functions.  EKS98$^{[6-7]}$ used E772 data with the
fitting done by eye only. The main improvements in EKPS$^{[8]}$ over
the EKS98 are the  automated  $\chi^2$ minimization, simplified and
better controllable fit functions, and most importantly, the
possibility for error estimates.  EKPS parametrization is found to
be fully consistent with the old EKS98 within the error estimates
obtained. It is noted that EKPS include the E866 and E772 data in
whole experimental kinematical range. The $\chi^2/d.o.f.$ is 0.916
$(84.3/92)$ with fitting the nuclear Drell-Yan data. It is obvious
that by reason of leaving the quark energy loss effect out, EKPS and
EKS98 overestimate the nuclear modification on sea quark
distributions. Therefore, the E866 collaboration gave a conclusion
that the energy loss effect can be negligible in nuclear Drell-Yan
reactions$^{[14]}$.  The HKN04$^{[10]}$ nuclear parton distributions
, which is extended to HKN07$^{[11]}$, added E772 and E866 Drell-Yan
data in the range $0.02< x_2<0.2$. It is considered that the
Drell-Yan cross section ratios is almost identical to the antiquark
ratio $\bar{q}^A(x_2)/\bar{q}^{A'}(x_2)$ in $x<0.1$ region. Our
calculation on the ratios of $R_{A/D}(x_{2})$ without quark energy
loss to those with linear quark energy loss reveals that the maximum
uncertainty from quark energy loss is approximately $3\%$ for nuclei
A(Ca,Fe)versus deuteron, and $6\%$ for W over D at $x_2\sim 0.1$.
Because the effects of  parton energy loss in the Drell-Yan process
are neglected in HKN analysis, the nuclear modification is yet
overestimated on sea quark distributions in Ca and Fe nuclei.

\vskip 0.5cm

{\bf 4  Summary }

\hspace{0.5cm} In this study, we have performed  a leading order
phenomenological analysis on nuclear Drell-Yan differential cross
section ratios as a function of the quark momentum fraction in the
beam proton and target nuclei for E772 experimental data. The quark
energy loss effect is quantified on the nuclear Drell-Yan
differential cross section ratios. The uncertainties due to quark
energy loss is discussed on determining nuclear sea quark
distribution from the proton-induced Drell-Yan reaction off nuclei.
With combining our previous works on the E866, it is concluded that
there is the quark energy loss effect apart from the nuclear effects
on the parton distribution as in deep inelastic scattering in
nuclear Drell-Yan process. The quark energy loss effect on cross
section ratios make greater with the increase of momentum fraction
of the target parton. It is found that the heavier the nucleus A,
the bigger the uncertainties due to quark energy loss.  The proton
incident middle and heavy nuclei Drell-Yan data result in  an
overestimate for nuclear modifications on sea quark distribution
functions if the quark energy loss effect is neglected. It is  hoped
that our results would provide  good directional information on the
magnitude and form of nuclear modifications on sea quark
distribution functions by means of the nuclear Drell-Yan
experimental data. Because of the large experimental error in E866
and E772 experimental data, the parameter $\alpha$ and  $\beta$ can
not be currently determined well. Therefore, we desire to operate
precise measurements of the experimental study from the relatively
low energy nuclear Drell-Yan process at J-PARC$^{[23]}$ and Fermilab
E906$^{[24]}$. These new experimental data on nuclear Drell-Yan
reaction can shed light on the energy loss of fast quark propagating
in a cold nuclei.

{\bf Acknowledgement:} This work is partially supported by Natural
Science Foundation of China(10575028) and  Natural Science
Foundation of Hebei Province(A2008000137).

\newpage

\begin{figure}
\centering
\includegraphics[width=1.0\textwidth]{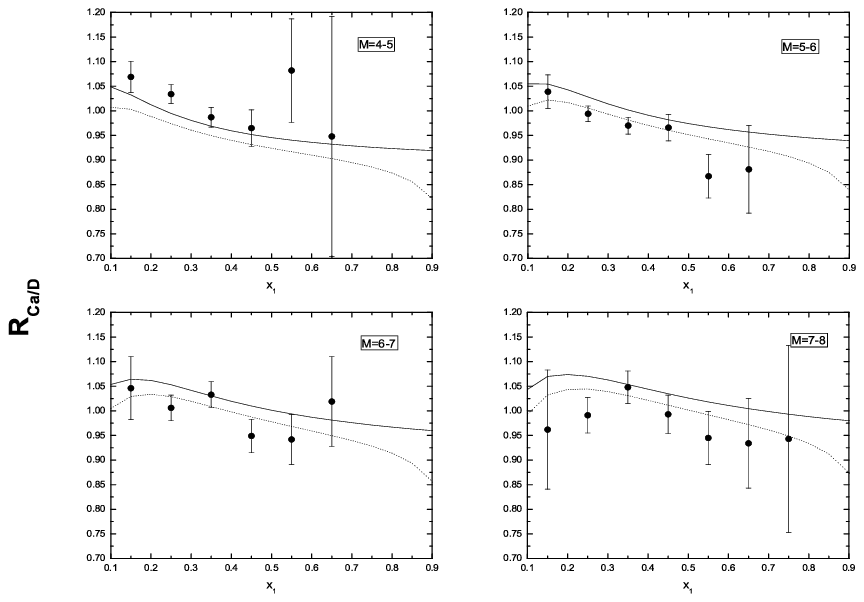}
\caption{ The nuclear Drell-Yan cross section ratios $R_{Ca/D}(x_1)$
 for various intervals $M$. Solid curves correspond to nuclear
effects on structure function. Dotted curves  show the combination
of linear quark energy loss effect with HKM cubic type of nuclear
parton distributions. The experimental data are taken from
E772$^{[14]}$.}
\end{figure}

\begin{figure}
\centering
\includegraphics[width=1.0\textwidth]{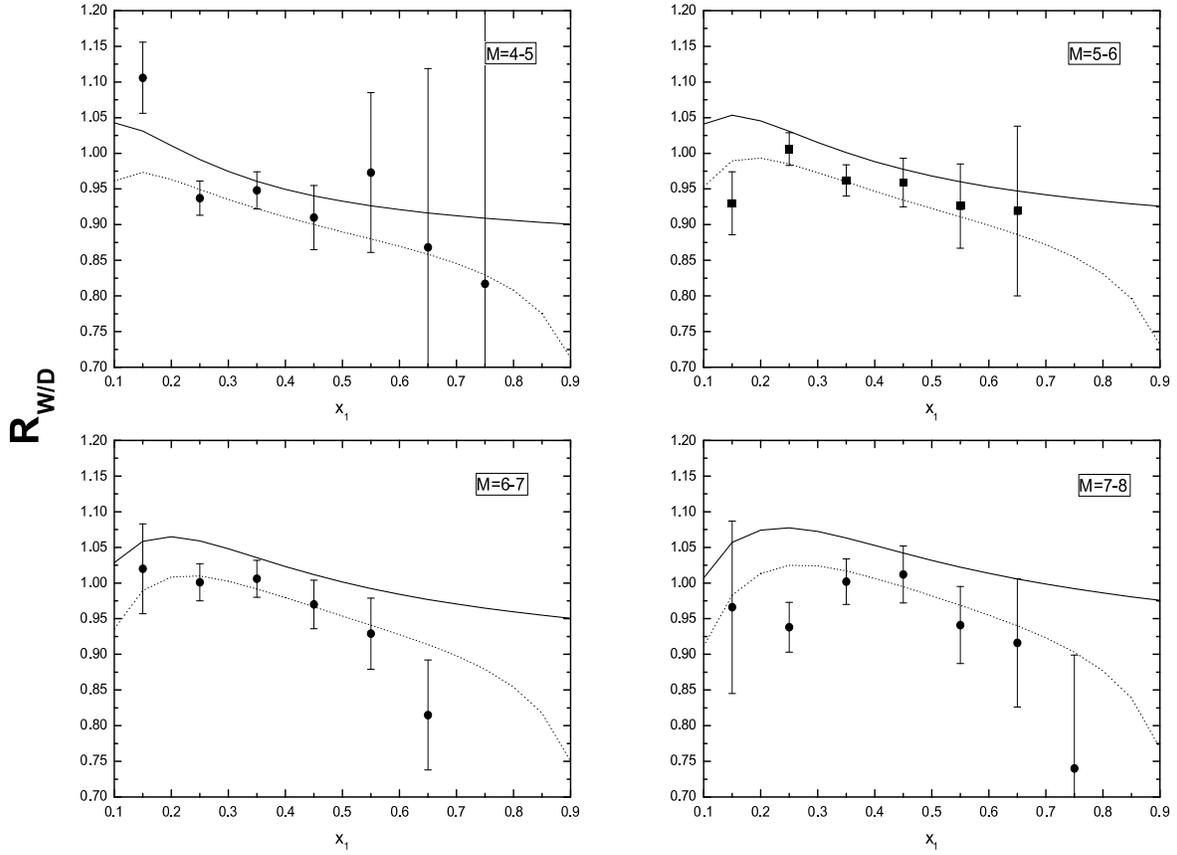}
\caption{The nuclear Drell-Yan cross section ratios $R_{W/D}(x_1)$
for various intervals $M$. The comments are the same as Fig.1}
\end{figure}

\begin{figure}
\centering
\includegraphics[width=1.0\textwidth]{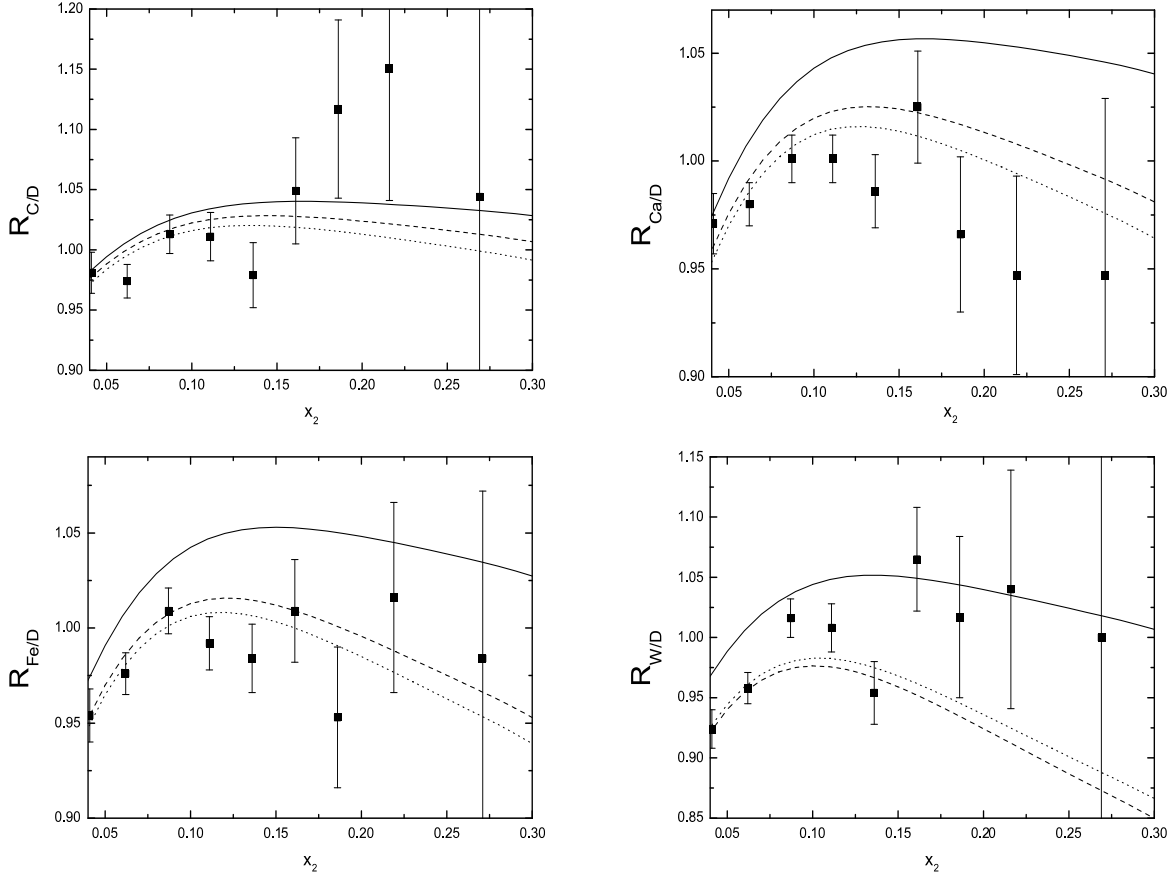}
\caption{The nuclear Drell-Yan cross section ratios $R_{A/D}(x_2)$
on nuclei(C,Ca,Fe,W) versus deuteron. Solid curves correspond to
nuclear effects on structure function. Dotted and dash curves show
the combination of HKM cubic type of nuclear parton distributions
with the quark energy loss $\alpha=1.26$ and $\beta=0.23$,
respectively. The experimental data are taken from  E772$^{[14]}$.}
\end{figure}

\end{document}